\documentclass[aps,prd,a4paper,nosuperscriptaddress,nofootinbib,showpacs,twocolumn,showkeys,amsfonts,amssymb,amsmath]{revtex4-2}
\usepackage{amssymb,latexsym}

\usepackage{bbold}
\usepackage{color}
\usepackage{amscd}
\usepackage{times}
\usepackage{epsfig}
\usepackage{psfrag}
\usepackage{graphicx}
\usepackage{amsthm,amsmath}
\usepackage{mathrsfs}
\usepackage[colorlinks=true, allcolors=blue]{hyperref}
\usepackage{leftindex}
\usepackage{mathtools}
\usepackage[dvipsnames]{xcolor}
\usepackage{physics}

\begin{document}
\title[]{Quantum gravitational contrast in creating Schr\"odinger cat state}

\author{Anupam Mazumdar}
\author{Tian Zhou}
\email{t.zhou@rug.nl}
\affiliation{ 
$^{1}$ Van Swinderen Institute, University of Groningen, 9747 AG Groningen, The Netherlands
}

\begin{abstract}
In this paper, we illustrate how a Schr\"odinger cat state created via a matter-wave interferometer can be viewed as the simplest quantum-gravity setup where we can treat both matter and gravity on an equal footing at a perturbative level. Here we treat Einstein's theory of general relativity using an effective field theory approach, quantising the massless spin-2 graviton in the presence of a quantum spatial superposition of matter that creates a matter-wave interferometer in the non-relativistic limit. We show that, due to the matter-graviton coupling, the graviton vacuum is displaced analogous to the coherent state. We study the contrast/overlap between the coherent states of the left and right superpositions in the matter-wave interferometer. We also study the entanglement between matter and the graviton in this setup and relate it to a gravitational contrast, or the overlap of the quantum geometries led by the coherent states. In the appendix, we provide an example of a time-dependent harmonic oscillator and study the contrast/overlap of such coherent states of the graviton.
\end{abstract}

\maketitle
\section{introduction}

A matter-wave interferometer is a growing topic of research from both theory and experimental perspective. Experimentally, it has been possible to create a matter wave interferometer from cold neutrons~\cite{Colella:1975,Werner:1979gi} to atomic masses~\cite{Fixler:2007is,Asenbaum:2016djh,schaff2014interferometry,Kovachi_2015,Asenbaum:2016djh,asenbaum2017phase}, and even heavier mesoscopic objects~\cite{Arndt_1999,Eibenberger:2013cqb,Haslinger:2014rxx,arndt2014testing,
Fein:2019dgf,pedalino2026probing,Margalit:2021}. While at low energies, the spacetime can be quantised by virtue of quantising the metric perturbations (the massless spin-2 graviton) in a covariant fashion around the Minkowski spacetime \cite{Gupta}, see also~\cite{Donoghue:1994dn} for an effective field theory treatment for gravity.
However, it is pertinent to note that a matter-wave interferometer can serve as an excellent tool for understanding the quantum nature of gravity at low energies through the quantum superposition of matter and the quantum nature of the gravitational interaction~\cite{Toros:2020krn,Rufo:2024ulr}. It was shown in the latter reference that quantum matter and the graviton are entangled in a tripartite system, e.g., in the matter-graviton vertex interaction.
In fact, this interplay between quantum superpositions of matter and the quantum nature of gravitational interaction yields quantum entanglement between two matter-wave interferometers, effectively entangling the Schr\"odinger cat states, induced by the graviton exchange~\cite{Bose:2017nin, ICTS}, which can be experimentally witnessed in a lab, see also~\cite{marletto2017gravitationally} and the experimental white paper~\cite{Bose:2025qns}. The entanglement induced by the virtual graviton exchange is possible {\emph{if and only if}} gravity is a bona fide quantum entity, as in the quantum gravity-induced entanglement of masses (QGEM) protocol, see~\cite{marshman2020locality,Bose:2022uxe,Vinckers:2023grv}, see also~\cite{
Belenchia:2018szb,
christodoulou2023locally,christodoulou2019possibility,Carney:2021vvt,Carney_2019,Danielson:2021egj}.  A couple of main applications for the QGEM protocol are that we can also probe massive graviton and the scale of new physics all the way to $10^{7}$~GeV, four orders of magnitude in physics beyond the Standard Model compared to the LHC. We can test the quantum version of the weak equivalence principle~\cite{Bose:2022czr,chakraborty2023distinguishing}, and test the light bending experiment in a lab, but via entanglement~\cite{Biswas:2022qto, Carney:2021vvt}.

The aim of this work is to show how a quantum spatial superposition of matter, which has a Gaussian wave function spread, induces a displaced vacuum for the gravitons in the spatial superposition, e.g. left and right arms of the matter-wave interferometer, thereby creating a coherent state with an occupation number of gravitons that depends on the infrared cut-off. It is not necessary to have the Gaussian spread of the wavefunction for each of the arms of the spatial superposition; however, if we had taken a Dirac delta distribution, the conclusion would remain the same, except that the displaced vacuum would yield the ultraviolet divergence. The displacement of the vacuum generates a finite contribution due to the Gaussian nature of the matter wavepacket. Furthermore, the combined state of matter and a graviton shows the superposition of spacetime geometry, whose wavefunction overlap gives a contrast in the geometry analogous to that in photon or matter wave interferometers, see ~\cite{Englert,Schwinger,Scully}. The contrast between the two arms, in fact, also determines the entanglement entropy between the matter and the graviton sector, which we will show in this paper. One crucial point, we will consider a closed system of matter-wave interferometer and the graviton sector. We will not take the decoherence effects in this paper, for the gravitational decoherence in the matter-wave system due to gravitational interaction can be found in many beautiful references, see \cite{Calzetta_1994,Anastopoulos:1995ya,Anastopoulos:2013zya,
Blencowe:2013,Ford:2015,Charles:2017,Danielson:2022tdw,Danielson:2022sga,Toros:2020krn,Parikh_2020, Toros:2020dbf,Biggs:2024dgp}. 
In all our computations, we assume the static regime. However, in the appendix, we will show that the time-dependent feature of a single harmonic oscillator that emits gravitational waves, see~\cite{Sabbata,Toros:2020krn}, will also appear in the computation of the overlap between the gravitational waves emitted by the quantum harmonic oscillator. After every oscillation, the gravitational waves are emitted, and we estimate the contrast of the gravitational waves upon every oscillation. In this context, the emission of gravitational waves leads to decoherence of the harmonic oscillator, as shown earlier in \cite{Toros:2020krn}.

In Section II, we discuss the displaced graviton vacuum arising from the coupling between quantum matter and the graviton in a perturbative quantum field theory of gravity. In Section III, we discuss the superposition of coherent graviton states in the presence of a quantum spatial superposition. In Section IV, we compute the overlap between the coherent states of the graviton. In section V, we conclude. In Appendix.\ref{appendixB}, we discuss the superposition of a propagating graviton and the overlap of the wave functions.

\section{Displaced gravitational vacuum}

In this section, we derive the Hamiltonian for a mass coupled with the gravitational field. First, in the weak-field approximation, the quantum gravitational field $g_{\mu\nu}$ is described as a quantised perturbation field $h_{\mu\nu}$ around the Minkowski background $\eta_{\mu\nu}$, namely~\footnote{We work in natural units, namely $\hbar=c=1$, and take the sign convention $(-,+,+,+)$ of the spacetime metric. 
},
\begin{equation}
    {g}_{\mu\nu}=\eta_{\mu\nu} + {h}_{\mu\nu}\, .
\end{equation}
We decompose the linear gravitational perturbation field ${h}_{\mu\nu}$ as \cite{Gupta}
\begin{equation}\label{decomposition}
{h}_{\mu\nu} = {\gamma}_{\mu\nu} -\frac{1}{2}\eta_{\mu\nu}{\gamma} \, ,
\end{equation}
where $\mathcal{A}\equiv\sqrt{16\pi G}$, and the field $\gamma_{\mu\nu}$ and $\gamma\equiv \eta^{\mu\nu}\gamma_{\mu\nu}$ represent the spin-2 and spin-0 mode, respectively, see~\cite{Biswas:2011ar,Biswas:2013kla}. Here our treatment follows that of Gupta's quantisation, where $\gamma_{\mu\nu}$ and $\gamma$ are treated as independent field quantities, and then the canonical quantisation of the two gravitational modes is given by\cite{Gupta}
\begin{equation}\label{quantistiongamma_munu}
\hat{\gamma}_{\mu\nu}=\mathcal{A}\int\frac{d^3\mathbf{k}}{(2\pi)^3} \frac{1}{\sqrt{2\omega_{\mathbf{k}}}}  \left(\hat{\mathcal{P}}_{\mu\nu}(\mathbf{k})e^{i\mathbf{k}\mathbf{x}} + \hat{\mathcal{P}}_{\mu\nu}^\dagger(\mathbf{k})e^{-i\mathbf{k}\mathbf{x}}\right)\, ,
\end{equation}
\begin{equation}\label{quantistiongamma}
\hat{\gamma} =2\mathcal{A}\int\frac{d^3\mathbf{k}}{(2\pi)^3} \frac{1}{\sqrt{2\omega_{\mathbf{k}}}}  \left(\hat{\mathcal{P}}(\mathbf{k})e^{i\mathbf{k}\mathbf{x}} + \hat{\mathcal{P}}^\dagger(\mathbf{k})e^{-i\mathbf{k}\mathbf{x}}\right) \, ,
\end{equation}
where $\omega_\mathbf{k}=|\mathbf{k}|$ and $\hat{\mathcal{P}}_{\mu\nu}$, $\hat{\mathcal{P}}_{\mu\nu}^\dagger$, $\hat{\mathcal{P}}$, and $\hat{\mathcal{P}}^\dagger$ are the annihilation and creation operators of spin-2 and spin-0 mode, respectively. The creation and annihilation operators obey the following commutation relations
~\cite{Gupta}:
\begin{equation}\label{commutator_p_munu}
[\hat{\mathcal{P}}_{\mu\nu}(\mathbf{k}),\hat{\mathcal{P}}_{\rho\sigma}^\dagger(\mathbf{k'})]=[\eta_{\mu\rho}\eta_{\nu\sigma}+\eta_{\mu\sigma}\eta_{\nu\rho}] (2\pi)^3\delta (\mathbf{k}-\mathbf{k}') \, ,
\end{equation}
\begin{equation}\label{commutator_p}
[\hat{\mathcal{P}}(\mathbf{k}),\ \hat{\mathcal{P}}^\dagger(\mathbf{k'})]= -(2\pi)^3\delta (\mathbf{k}-\mathbf{k}')\, .
\end{equation}
The Hamiltonian of gravitons in the Minkowski background is given by~\cite{Gupta}:
\begin{equation}
   {\hat{H}}_g=\int\frac{d\mathbf{k}}{(2\pi)^3} \, \omega_\mathbf{k}  \left( \frac{1}{2} \hat{\mathcal{P}}^\dagger_{\mu\nu}(\mathbf{k})\hat{\mathcal{P}}^{\mu\nu}(\mathbf{k})  - \hat{\mathcal{P}}^\dagger(\mathbf{k})\hat{\mathcal{P}}(\mathbf{k}) \right)\, .
\end{equation}
Considering the minimal coupling between matter and the gravitational field, the interaction Hamiltonian in the perturbative limit is given by:
\begin{align}\label{Hint}
    {\hat{H}}_{int}&=-\frac{1}{2} \int d\mathbf{x}\ {\hat{h}}_{\mu\nu}(\mathbf{x}) {\hat{T}}^{\mu\nu}(\mathbf{x})\nonumber\\
    &=\frac{1}{2}\int d\mathbf{x}\ \left[ {\hat{\gamma}}_{\mu\nu}(\mathbf{x})+\frac{1}{2}\eta_{\mu\nu}\hat{\gamma} (\mathbf{x})\right] {\hat{T}}^{\mu\nu}(\mathbf{x})\, ,
\end{align}
where ${\hat{T}}_{\mu\nu}(\mathbf{x})$ is the energy-momentum tensor operator of matter. In momentum space, the interaction Hamiltonian can be expanded as
\begin{align}\label{Hint}
   {\hat{H}}_{int}=&\int\frac{d^3\mathbf{k}}{(2\pi)^3} \frac{\mathcal{A}}{2}  \frac{1}{\sqrt{2\omega_{\mathbf{k}}}}  \left[\hat{\mathcal{P}}_{\mu\nu}(\mathbf{k})-\eta_{\mu\nu}\hat{\mathcal{P}}(\mathbf{k})\right]{\hat{T}}^{\mu\nu}(\mathbf{k})\nonumber\\
    & + h.c.  \, ,
\end{align}
where $h.c.$ represents the Hermitian conjugate, and ${T}_{\mu\nu}(\mathbf{k})$ is the Fourier transformation of the energy-momentum tensor
\begin{equation}
    {\hat{T}}^{\mu\nu}(\mathbf{k})\equiv \int d\mathbf{x}\, {\hat{T}}^{\mu\nu}(\mathbf{x}) e^{-i\mathbf{k}\cdot\mathbf{x}}\, .
\end{equation}
Due to the interaction Hamiltonian, the gravitational vacuum state of the total Hamiltonian ${\hat{H}}=\hat{H}_g+\hat{H}_{int}$, denoted by $\ket{\Omega}$, is a displaced vacuum, defined by
\begin{equation}\label{Omega}
    \ket{\Omega} \equiv \hat{D}(\hat{\alpha}_{\mu\nu},\hat{\alpha})\ket{0} \, , 
\end{equation}
where $\ket{0}$ denotes the vacuum state of ${H}_g$, and the displacement operator $D(\hat{\alpha}_{\mu\nu},\hat{\alpha})$ is defined by
\footnote{Note that we define the displacement parameters as operator-valued quantities because they are made up with the energy-momentum operator $\hat{T}_{\mu\nu}$ of matter.}
\begin{align}
    \hat{D} =&\exp [\int \frac{d\mathbf{k}}{(2\pi)^3}( \hat{\alpha}^{\mu\nu}(\mathbf{k})\hat{\mathcal{P}}_{\mu\nu}^\dagger(\mathbf{k}) - \hat{\alpha}^{\mu\nu*}(\mathbf{k}) \hat{\mathcal{P}}_{\mu\nu}(\mathbf{k}) )] \nonumber \\
    &\times\exp [\int \frac{d\mathbf{k}}{(2\pi)^3}( \hat{\alpha}(\mathbf{k})\hat{\mathcal{P}}^\dagger(\mathbf{k}) - \hat{\alpha}^*(\mathbf{k}) \hat{\mathcal{P}}(\mathbf{k}))]\, .
\end{align}
The displacement parameter $\hat{\alpha}_{\mu\nu}$ and $\hat{\alpha}$ is solved from the equation
\begin{equation}\label{DH0D}
    \hat{D}(\hat{\alpha}_{\mu\nu},\hat{\alpha})\hat{H}_g \hat{D}^\dagger(\hat{\alpha}_{\mu\nu},\hat{\alpha}) = \hat{H}_g + \hat{H}_{int} + \hat{\beta} \, ,
\end{equation}
where $\hat{\beta}$ is an operator composed solely of the energy-momentum tensor of matter. According to Eq.(\ref{Hint},\ref{DH0D},\ref{DP00P00D},\ref{DPPD}), we obtain
\begin{align}
    \hat{\alpha}_{\mu\nu}(\mathbf{k})&=-\sqrt{\frac{2\pi G}{\omega_\mathbf{k}^3}}\, \hat{T}^\dagger_{\mu\nu}(\mathbf{k}) \, , \label{alphamunuk}\\
    \hat{\alpha}(\mathbf{k})&= \sqrt{\frac{2\pi G}{\omega_\mathbf{k}^3}} \eta^{\mu\nu} \hat{T}_{\mu\nu}^\dagger (\mathbf{k}) \, , \label{alphak}
\end{align}
and the constant term $\hat{\beta}$ reads
\begin{align}\label{beta}
    \hat{\beta} &=\int \frac{d\mathbf{k}}{(2\pi)^3}\ \omega_{\mathbf{k}}(2|\hat{\alpha}_{\mu\nu}(\mathbf{k})|^2 - |\hat{\alpha}(\mathbf{k})|^2 ) \nonumber\\
    &= 2\pi G\int \frac{d\mathbf{k}}{(2\pi)^3} \frac{2\hat{T}^\dagger_{\mu\nu}(\mathbf{k})\hat{T}^{\mu\nu}(\mathbf{k})-|\hat{T}_{\mu}^{\mu}(\mathbf{k})|^2}{\omega_\mathbf{k}^2} \, .
\end{align}
The $\hat{\beta}$ term is related to the energy shift of the gravitational vacuum due to the existence of matter, namely $\hat{H}\ket{\Omega}=\hat{D}\hat{H}_0\hat{D}^\dagger\ket{\Omega}-\hat{\beta}\ket{\Omega}=-\hat{\beta}\ket{\Omega}$. The formula \eqref{beta} is consistent with the gravitational energy shift obtained by the second-order perturbation theory\cite{Bose:2022uxe}.

The number operator of gravitons is defined by
\begin{equation}
    \hat{N}\equiv \int\frac{d\mathbf{k}}{(2\pi)^3} \,  \left( \frac{1}{2} \hat{\mathcal{P}}^\dagger_{\mu\nu}(\mathbf{k})\hat{\mathcal{P}}^{\mu\nu}(\mathbf{k})  - \hat{\mathcal{P}}^\dagger(\mathbf{k})\hat{\mathcal{P}}(\mathbf{k}) \right)\, .
\end{equation}
For the displaced vacuum, the number of gravitons is given by
\begin{align}
    \hat{N}=&\int \frac{d\mathbf{k}}{(2\pi)^3}\ (2|\hat{\alpha}_{\mu\nu}(\mathbf{k})|^2 - |\hat{\alpha}(\mathbf{k})|^2 )\nonumber\\
    =&2\pi G\int \frac{d\mathbf{k}}{(2\pi)^3} \frac{2\hat{T}^\dagger_{\mu\nu}(\mathbf{k})\hat{T}^{\mu\nu}(\mathbf{k})-|\hat{T}_{\mu}^{\mu}(\mathbf{k})|^2}{\omega_\mathbf{k}^3} \, .
\end{align}
In addition, the expectation value of the field operators ${\hat{\gamma}}_{\mu\nu}(\mathbf{x})$ and ${\hat{\gamma}} (\mathbf{x})$ of the displaced vacuum (\ref{Omega},\ref{alphamunuk},\ref{alphak}) are given by
\begin{align}\label{EVgamma_munu}
    &\bra{\Omega}{\hat{\gamma}}_{\mu\nu}(\mathbf{x})\ket{\Omega} = \bra{0} \hat{D}^\dagger {\hat{\gamma}}_{\mu\nu}(\mathbf{x}) \hat{D}\ket{0} \nonumber\\
    &= - 8\pi G \int\frac{d\mathbf{k}}{(2\pi)^3} \frac{\hat{T}_{\mu\nu}(\mathbf{k})}{\omega_{\mathbf{k}}^2}  e^{i\mathbf{k}\mathbf{x}} + h.c. \, ,
\end{align}
\begin{equation}\label{EVgamma}
    \bra{\Omega}{\hat{\gamma}}(\mathbf{x})\ket{\Omega} =  8\pi G \int\frac{d\mathbf{k}}{(2\pi)^3} \frac{\hat{T}_\mu^\mu(\mathbf{k})}{\omega_{\mathbf{k}}^2}  e^{i\mathbf{k}\mathbf{x}} + h.c.
\end{equation}
where we have applied the formula \eqref{Dgamma00D} and \eqref{DgammaD} in Appendix. \ref{appendixA}.

Let us consider an energy-momentum tensor with a Gaussian mass density distribution, namely
\footnote{Here we choose a Gaussian distribution of matter to regulate the UV behaviour. However, there are other regularizations as well, such as uniform spherical mass distributions, see Ref. \cite{diosi2013gravity,howl2019exploring}.  }
\begin{equation}
    \hat{T}^{\mu\nu}(\mathbf{x})= \frac{p^\mu p^\nu}{E}{(2\pi \sigma^2)^{-\frac{3}{2}}}\exp \left(-\frac{(\mathbf{x}-\hat{\mathbf{x}}_c)^2}{2\sigma^2} \right) \, ,
\end{equation}
where $p^{\mu}=(E,p,0,0)$ represents the four-momentum of the mass, $\hat{\mathbf{x}}_c$ is the location operator of the centre of mass, and $\sigma$ represents the width of the mass density distribution. In momentum space, the energy-momentum tensor is given by
\begin{equation}\label{T00static}
    \hat{T}^{\mu\nu}(\mathbf{k})= \frac{p^\mu p^\nu}{E} \exp \left( -i\mathbf{k}\cdot \hat{\mathbf{x}}_c - \frac{1}{2}\sigma^2 |\mathbf{k}|^2  \right)\, .
\end{equation}
Employing the energy-momentum tensor \eqref{T00static}, we get the displacement parameters (\ref{alphamunuk},\ref{alphak}) and the energy-shift (\ref{beta}) due to the existence of the matter
\begin{equation}
\hat{\beta} = 2\pi G\int \frac{d\mathbf{k}}{(2\pi)^3} \frac{M^2 e^{-\sigma^2|\mathbf{k}|^2}}{\omega_\mathbf{k}^2}=\frac{GM^2}{2\sqrt{\pi}\sigma}\, .
\end{equation}
where $M=\sqrt{E^2-p^2}$ is the static mass of the matter. The operator $\hat{\beta}$ turns into a constant, which is only determined by the total mass and its distribution. In the limit $\sigma \rightarrow 0$, in the limit of the Dirac Delta distribution of matter, the energy shift in the displaced vacuum will blow up; it is the standard ultraviolet divergence in general relativity. Similarly, the number of gravitons can be made ultraviolet finite in the Gaussian density profile, but now the infrared part is divergent, and is given by
\begin{align}
    N&=\frac{GM^4}{\pi E^2}\int^\infty_\epsilon dk  \frac{e^{-\sigma^2k^2}}{k} \nonumber\\
    &\approx \frac{GM^4}{\pi E^2}\left[-\ln(\sigma\epsilon)- \frac{\gamma_E}{2} + \frac{\sigma^2\epsilon^2}{2} \right] \, ,
\end{align}
where we have set an infrared cut-off $\epsilon$ for the integral and $\gamma_E$ is the Euler constant. The occupation number of gravitons by the displaced vacuum is determined by the ultraviolet and infrared cut-off scales of gravitons.


\section{Superposition of coherent states of graviton and geometry}

In this section, we consider a massive particle as a static quantum spatial superposition. Here, we take the local mass approximation, namely
\begin{equation}
\hat{\mathbf{x}}_c \ket{\phi_L} \approx \mathbf{x}_L \ket{\phi_L} \,,~\hat{\mathbf{x}}_c \ket{\phi_R} \approx \mathbf{x}_R\ket{\phi_R} \, ,
\end{equation}
where $\mathbf{x}_{L, R}$ label the position of the left and right branch of the spatial superposition, respectively. The momentum of the mass is taken as 
\begin{equation}\label{momentum}
p^{\mu}_L= p^{\mu}_R=(M,0,0,0)^T \, .
\end{equation}
Now, taking the displaced gravitational vacuum state into account, the complete quantum state of the system (joint graviton and matter) can be written as
\begin{equation}
\ket{\Psi} = \frac{1}{\sqrt{2}}( \ket{\phi_L} \ket{\Omega_L} + \ket{\phi_R} \ket{\Omega_R} )\, ,
\end{equation}
where $\ket{\phi_L}$ and $\ket{\phi_R}$ represent the two branches of the spatial superposition state of the mass. Therefore, the two branches of the coherent states of the graviton are given by:
\begin{equation}\label{OmegaLR}
    \ket{\Omega_L} =\hat{D}(\alpha^{\mu\nu}_L,\alpha_L)\ket{0}\, ,\ \ket{\Omega_R} =\hat{D}(\alpha^{\mu\nu}_R,\alpha_R)\ket{0}\, ,
\end{equation}
where the displaced parameters are given by
\begin{equation}\label{alphamunuL}
    \alpha^{\mu\nu}_L(\mathbf{k})=-\sqrt{\frac{2\pi G }{\omega_\mathbf{k}^3}} \frac{p_L^\mu p_L^\nu}{E}\, e^{i\mathbf{k}\cdot\mathbf{x}_L-\sigma^2|\mathbf{k}|^2/2}\, ,
\end{equation}
\begin{equation}\label{alphaL}
\alpha_L(\mathbf{k})=-\sqrt{\frac{2\pi G }{\omega_\mathbf{k}^3}} \frac{M^2}{E}\, e^{i\mathbf{k}\cdot\mathbf{x}_L-\sigma^2|\mathbf{k}|^2/2}\, ,
\end{equation}
\begin{equation}\label{alphamunuR}
    \alpha^{\mu\nu}_R(\mathbf{k})=-\sqrt{\frac{2\pi G }{\omega_\mathbf{k}^3}} \frac{p_R^\mu p_R^\nu}{E}\, e^{i\mathbf{k}\cdot\mathbf{x}_R-\sigma^2|\mathbf{k}|^2/2}\, ,
\end{equation}
\begin{equation}\label{alphaR}
\alpha_R(\mathbf{k})=-\sqrt{\frac{2\pi G }{\omega_\mathbf{k}^3}} \frac{M^2}{E}\, e^{i\mathbf{k}\cdot\mathbf{x}_R-\sigma^2|\mathbf{k}|^2/2}\, .
\end{equation}
The spatial superposition of mass is entangled with the superposition state of coherent states of gravitons. 

We now demonstrate that the displaced vacuum $\ket{\Omega_{L, R}}$ corresponds to two Schwarzschild geometries centered at $\mathbf{x}_L$ and $\mathbf{x}_R$, respectively. Let us focus on the expectation value of the spin-2 and spin-0 fields with respect to the displaced vacuum $\ket{\Omega_L}$. Applying Eq. (\ref{EVgamma_munu},\ref{EVgamma}), we have
\begin{equation}
   \langle{\hat{\gamma}}_{\mu\nu}({r})\rangle_L=-\frac{4GM}{{r}}\frac{(p_L)_\mu (p_L)_\nu}{E}\erf \left( \frac{{r}}{\sqrt{2}\sigma} \right) \, ,
\end{equation}
\begin{equation}
    \langle{\hat{\gamma}}({r})\rangle_L= -\frac{4GM}{{r}}\erf \left( \frac{{r}}{\sqrt{2}\sigma} \right) \, ,
\end{equation}
where ${r}\equiv |\mathbf{x}-\mathbf{x}_L|$ is the distance to position $\mathbf{x}_L$, $\langle \hat{\mathcal{O}} \rangle_L\equiv \bra{\Omega_L} \hat{\mathcal{O}} \ket{\Omega_L}$ denotes the expectation value of operators with respect to $\ket{\Omega_L}$, and $\erf(x)$ denotes the error function. Therefore, the expectation value of the gravitational field $\hat{h}_{\mu\nu}$ with respect to the displaced vacuum $|\Omega_L\rangle$ is given by
\begin{align}\label{EVh_munu}
    \langle{\hat{h}}_{\mu\nu}({r})\rangle_L &= \langle {\hat{\gamma}}_{\mu\nu}({r})\rangle_L-\frac{\eta_{\mu\nu}}{2}\langle{\hat{\gamma}}({r}) \rangle_L \, \nonumber\\
    &=  -\frac{2G}{{r}}\frac{2(p_L)_\mu (p_L)_\nu-\eta_{\mu\nu}M^2}{E} \erf \left( \frac{{r}}{\sqrt{2}\sigma} \right)\, .
\end{align}
Applying the momentum \eqref{momentum} for a static spatial superposition, it is clear to see that $ \langle{h}_{00}({r})\rangle= -\langle{h}_{11}({r})\rangle  \approx -2GM/{r}$ tends to Schwarzchild geometry\footnote{In Ref. \cite{giddings2025quantum,donnelly2016diffeomorphism}, the authors constructed other gravitational configurations of a particle based on the gauge-invariance condition, these configurations have higher or even more singular profile compared to the Schwarzchild geometry in the linearized limit \cite{donnelly2016diffeomorphism}. In our case, the gravitational ground state is the displaced vacuum, see Eq.~\eqref{Omega}, which creates our solution Eq.~(\ref{EVh_munu}), which is non-singular by virtue of the Gaussian profile of the density.} in the large-distance limit (also in the point-like source limit), namely $ r \gg \sigma$. 

Further, we derive the quantum operators and the expectation values of the components of the Riemann tensor with respect to the displaced vacuum $\ket{\Omega_L}$. According to the definition of the gravitational perturbation field, the linear Riemann tensor is given by
\begin{align}
    \hat{R}_{\mu\nu\rho\sigma}=&\frac{1}{2}(\partial_\nu\partial_\rho \hat{h}_{\mu\sigma}+\partial_\mu\partial_\sigma \hat{h}_{\nu\rho}-\partial_\mu\partial_\rho \hat{h}_{\nu\sigma}-\partial_\nu\partial_\sigma \hat{h}_{\mu\rho}) \nonumber\\
    =&\frac{1}{2}(\partial_\nu\partial_\rho {\hat{\gamma}}_{\mu\sigma}+\partial_\mu\partial_\sigma {\hat{\gamma}}_{\nu\rho}-\partial_\mu\partial_\rho {\hat{\gamma}}_{\nu\sigma}-\partial_\nu\partial_\sigma {\hat{\gamma}}_{\mu\rho}) \nonumber\\
    & -\frac{1}{4} (\eta_{\mu\sigma}\partial_\nu\partial_\rho {\hat{\gamma}}+\eta_{\nu\rho}\partial_\mu\partial_\sigma {\hat{\gamma}}-\eta_{\nu\sigma}\partial_\mu\partial_\rho {\hat{\gamma}}\nonumber\\
    &-\eta_{\mu\rho}\partial_\nu\partial_\sigma {\hat{\gamma}}) \, .
\end{align}
In the spherical coordinate $(t,r,\theta,\phi)$, we have
\begin{equation}
    \hat{R}_{trtr} = -\frac{1}{2}  \left( \partial_r^2  {\hat{\gamma}}_{00}  + \frac{1}{2} \partial_r^2{\hat{\gamma}}\right) \, .
\end{equation}
Therefore, we obtain the expectation value of components of the Riemann tensor
\begin{eqnarray}
\label{sch}
    \langle \hat{R}_{trtr} \rangle_L 
    &=& -\frac{1}{2} \partial_r^2 \bra{\Omega_L} {\hat{\gamma}}_{00}+\frac{{\hat{\gamma}}}{2}\ket{\Omega_L}\nonumber \\ 
    &=&\frac{1}{2} \frac{\partial^2}{\partial r^2} \frac{2GM}{r}{\rm erf} \left( \frac{r}{\sqrt{2}\sigma}\right)\nonumber \\    
   & \xrightarrow{ r \gg \sigma} &\frac{2GM}{r^3} \,.
\end{eqnarray}
For the right branch of the displaced vacuum, namely $\ket{\Omega_R}$, the expectation value $\langle{\hat{h}}_{\mu\nu}\rangle_R$ and $ \langle \hat{R}_{trtr} \rangle_R$ takes the same form of Eq. \eqref{EVh_munu} and Eq. \eqref{sch}, in which the only difference is that the distance $r$ is respect to position $\mathbf{x}_R$ in this case. It shows that the static spatial superposition of matter produces the superposition of the Schwarzschild spacetimes centred at two different positions.

To show the entanglement between mass and gravitational field, we trace out the gravitational part of the total density matrix $\rho\equiv\ket{\Psi}\bra{\Psi}$. Therefore, we get the partial density matrix of mass
\begin{align}
\rho_m = \frac{1}{2}(&\ket{\phi_L}\bra{\phi_L}+ C \ket{\phi_R}\bra{\phi_L} \nonumber\\
&+ C^*\ket{\phi_L}\bra{\phi_R}+\ket{\phi_R}\bra{\phi_R})
\end{align}
where $C$ denotes the overlap between the two branches of the gravitational vacuum, which can be thought of as a quantum bridge between the superposition of geometries led by the coherent states of the graviton vacuum, e.g. corresponding to the left and right components of the matter wave interferometer. 
\begin{equation}
    C\equiv  \langle \Omega_L| \Omega_R \rangle\, .
\end{equation}
The entanglement entropy between matter and the gravitational field is therefore given by
\begin{align}\label{entropy}
S&= -{\rm Tr} (\rho_m \ln \rho_m) \nonumber\\
&=-\frac{1-C}{2}\ln \left(\frac{1-C}{2}\right)-\frac{1+C}{2}\ln \left(\frac{1+C}{2}\right) \, .
\end{align}
In the case of $C=1$, the total quantum state is separable, so there is no entanglement. In the case of $C\rightarrow0$, the two branches of the gravitational vacuum are almost orthogonal, which causes the maximal entanglement state with entropy $S_{max}=\ln 2$.

\section{Contrast in the coherent states of the graviton}
\begin{figure}
    \centering
    \includegraphics[width=0.8\linewidth]{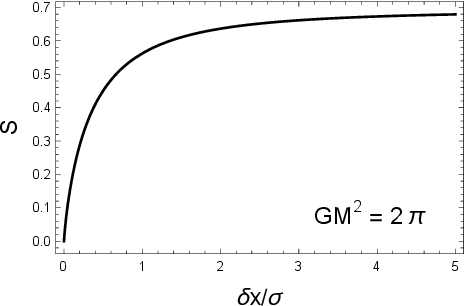}
    \includegraphics[width=0.8\linewidth]{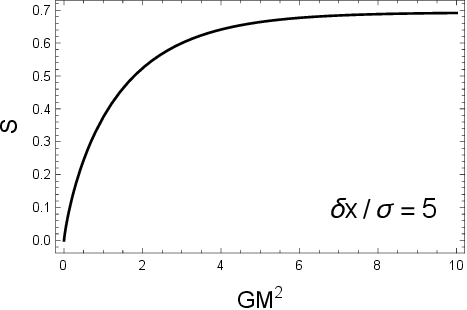}
    \caption{The entanglement entropy between the matter and the coherent state of the graviton as a function of the superposition size $\delta x$ and the mass $M$ of the matter, where we set $GM^2=2\pi$ and $\delta x/\sigma=5$, respectively.}
    \label{integral}
\end{figure}

From Eqs.~(\ref{OmegaLR},\ref{alphamunuL},\ref{alphaL},\ref{alphamunuR},\ref{alphaR}), the overlap of the coherent state of the graviton  superposition can be evaluated by
\begin{align}\label{overlap1}
C=&\bra{0} D^\dagger(\alpha^{\mu\nu}_L,\alpha_L) D(\alpha^{\mu\nu}_R,\alpha_R) \ket{0}  \nonumber\\
& \exp\left(-\int \frac{d\mathbf{k}}{(2\pi)^3} |\alpha^{\mu\nu}_{L}(\mathbf{k})-\alpha^{\mu\nu}_{R}(\mathbf{k})|^2 \right)\nonumber\\
&\times\exp\left(\int \frac{d\mathbf{k}}{(2\pi)^3} \frac{1}{2}|\alpha_L(\mathbf{k})-\alpha_R(\mathbf{k})|^2 \right) \nonumber\\
=& \exp\left(-\int \frac{d\mathbf{k}}{(2\pi)^2} \frac{GM^2 e^{-\sigma^2|\mathbf{k}|^2}}{\omega_{\mathbf{k}}^3}  [1-\cos(\mathbf{k}\cdot\delta\mathbf{x})] \right) \nonumber\\
=& \left( 1+ \frac{\delta x}{4\sigma}\right)^{-\frac{GM^2}{2\pi}} \, ,
\end{align}
where $\delta \mathbf{x}\equiv \mathbf{x}_L-\mathbf{x}_R$ and $\delta x\equiv |\delta\mathbf{x}|$ is the spatial superposition size. As we can see, for a large mass $M$, but finite $\delta x/\sigma$, $C\rightarrow 0$, showing maximum entanglement.  This is a very interesting result, showing that a massive object can be maximally entangled with the coherent state of the graviton. When $\sigma \rightarrow 0$, again, for a finite $M$, we get $C\rightarrow 0$, and the system of matter and graviton states is entangled. 
The entanglement entropy between the matter and the quantum spacetime geometry is shown in Fig. \ref{integral}. 
A similar analysis can be performed in a time-dependent case, where, for illustration, we considered a quantum harmonic oscillator in a Gaussian state that emits gravitational waves, see~\cite{Sabbata,Toros:2020krn}. We can ask a similar question about the overlap of the coherent state of the gravitons emitted during the oscillations. We have shown this computation in Appendix. \ref{appendixB}.


\section{Discussions}

In this paper, we have shown that when we treat matter and the graviton on an equal footing, the graviton vacuum evolves to a coherent state. The displaced vacuum yields a classical geometry, where the individual, left, and right half of the matter superposition yields a geometry equivalent to a superposition of a Schwarzschild geometry, as shown in Eq.~(\ref{sch}) in a perturbative treatment of gravity.

Furthermore, we have shown the quantum overlap, which we may call a {\it quantum bridge}, between the two coherent states of the graviton changes with mass, superposition size, and the width of the individual mass density profile. In this analysis, we ignored the time-dependent aspects; however, these can be added in the future, and there will also be a dependence on the momentum of the individual wave packets that create the matter-wave superposition.

In Appendix. \ref{appendixB}, we present a similar computation in a time-dependent setting for a quantum harmonic oscillator. Such a system emits gravitational waves, and we computed the overlap of the displaced vacuum of the gravitons in the transverse-traceless gauge. 

The results of the paper shed light on how matter-graviton coupling provides the quantum overlap between quantum geometries lead by the displaced vacuum of the gravitons at a perturbative level in a spatial superposition of matter. The results corroborate the results of the papers that show how to witness this quantum superposition of geometry via entanglement witness~\cite{Bose:2017nin}. Note that we cannot conjecture beyond the tree level, as higher order terms, such as post-Newtonian contributions and higher order vertex corrections will need to be taken into account to go beyond the computations we have performed in this paper. At a perturbative level our computations of the overlap/contrast between the geometries during quantum spatial superposition provide a hint how quantum gravity may pave out in reality when the non-perturbative effects are taken into account, which are indeed hard questions and demand a non-perturbative treatment of gravity.

{\it Acknowledgements}: 
A.M.’s research is funded
by the Gordon and Betty Moore Foundation through Grant GBMF12328, DOI 10.37807/GBMF12328. This material is
based on work supported by the Alfred P. Sloan Foundation under Grant No. G-2023-21130. T.Z. is supported by the China Scholarship Council (CSC).

\appendix

\section{Equations}\label{appendixA}

Based on the commutation relation \eqref{commutator_p_munu} and \eqref{commutator_p}, one can obtain the following equations
\begin{equation}
    \hat{D}({\alpha}_{\mu\nu},\alpha) \hat{\mathcal{P}}_{\rho\sigma}\hat{D}^\dagger({\alpha}_{\mu\nu},\alpha)= \hat{\mathcal{P}}_{\rho\sigma}-2\alpha_{\rho\sigma}\, , 
\end{equation}
\begin{equation}
     \hat{D}({\alpha}_{\mu\nu},\alpha) \hat{\mathcal{P}}_{\rho\sigma}^\dagger \hat{D}^\dagger({\alpha}_{\mu\nu},\alpha)= \hat{\mathcal{P}}_{\rho\sigma}-2\alpha_{\rho\sigma}^*\, ,
\end{equation}
\begin{equation}
    \hat{D}({\alpha}_{\mu\nu},\alpha) \hat{\mathcal{P}}\hat{D}^\dagger({\alpha}_{\mu\nu},\alpha)= \hat{\mathcal{P}}+\alpha\, ,
\end{equation}
\begin{equation}
    \hat{D}({\alpha}_{\mu\nu},\alpha) \hat{\mathcal{P}}^\dagger \hat{D}^\dagger({\alpha}_{\mu\nu},\alpha)= \hat{\mathcal{P}}^\dagger+\alpha^* \, .
\end{equation}
Then, we have
\begin{align}\label{DP00P00D}
    \hat{D} \hat{\mathcal{P}}_{\mu\nu}^\dagger\hat{\mathcal{P}}^{\mu\nu}\hat{D}^\dagger =& \hat{\mathcal{P}}_{\mu\nu}^\dagger\hat{\mathcal{P}}^{\mu\nu} - 2({\alpha}_{\mu\nu}^* \hat{\mathcal{P}}^{\mu\nu} + \alpha^{\mu\nu} \hat{\mathcal{P}}_{\mu\nu}^\dagger) \nonumber\\
    &+ 4 |{\alpha}_{\mu\nu}|^2 \, ,
\end{align}
\begin{equation}\label{DPPD}
    \hat{D} \hat{\mathcal{P}}^\dagger\hat{\mathcal{P}} \hat{D}^\dagger= \hat{\mathcal{P}}^\dagger\hat{\mathcal{P}} + (\alpha^* \hat{\mathcal{P}} + \alpha \hat{\mathcal{P}}^\dagger) +  |\alpha|^2 \, ,
\end{equation}

\begin{align}\label{Dgamma00D}
    \hat{D}{\hat{\gamma}}_{00}(\mathbf{x}) \hat{D}^\dagger=&{\hat{\gamma}}_{00}(\mathbf{x}) - \int\frac{d\mathbf{k}}{(2\pi)^3} \frac{2\mathcal{A}}{\sqrt{2\omega_{\mathbf{k}}}}  [\alpha_{00}(\mathbf{k})e^{i\mathbf{k}\mathbf{x}} \nonumber\\
    &+ \alpha^*_{00}(\mathbf{k})e^{-i\mathbf{k}\mathbf{x}}] 
\end{align}

\begin{align}\label{DgammaD}
    \hat{D}{\hat{\gamma}}(\mathbf{x})\hat{D}^\dagger=&{\hat{\gamma}}(\mathbf{x}) + \int\frac{d\mathbf{k}}{(2\pi)^3} \frac{2\mathcal{A}}{\sqrt{2\omega_{\mathbf{k}}}}  [\alpha(\mathbf{k})e^{i\mathbf{k}\mathbf{x}}+ \alpha^*(\mathbf{k})e^{-i\mathbf{k}\mathbf{x}}] \, .
\end{align}

\section{Overlap of the displaced gravitons in a time-dependent setup}\label{appendixB}

In this section, we focus on the quantum dynamics of gravitational waves and the superposition state of propagating gravitons. In the interaction Hamiltonian \eqref{Hint}, we quantize the $\{i,j\}\in\{1,2,3\}$ components of the gravitational perturbation operator in the transverse traceless(TT) gauge, namely\cite{bose2022infrared, bose2021gravitons,Toros:2020krn}
\begin{equation}
\hat{h}^{TT}_{ij}(\mathbf{x})=\sum_\lambda \int \frac{d\mathbf{k}}{(2\pi)^3} \frac{\mathcal{A}}{\sqrt{2\omega_\mathbf{k}}}\, {\rm e}_{ij}^\lambda (\mathbf{n}) \hat{g}_{\mathbf{k},\lambda}e^{i\mathbf{k}\cdot \mathbf{x}} + {\rm H.c.} \, ,
\end{equation}
where $\hat{g}_{\mathbf{k},\lambda}$ is the graviton annihilation operator, $e^{ij}_{\lambda}$ denotes the two polarization basis of the GW modes, $\lambda=1,2$, $\mathbf{n}\equiv \mathbf{k}/|\mathbf{k}|$. The Hamiltonian of gravitational waves is given by
\begin{equation}
    \hat{H}_g=\sum_\lambda\int d\mathbf{k} \, \omega_\mathbf{k} \hat{g}^\dagger_{\mathbf{k},\lambda}\hat{g}_{\mathbf{k},\lambda} \, .
\end{equation}
where the creation and annihilation operator satisfies the commutation relation $[\hat{g}_{\mathbf{k},\lambda}, \hat{g}^\dagger_{\mathbf{k}',\lambda'}]=\delta(\mathbf{k}-\mathbf{k}')\delta_{\lambda\lambda'}$. Considering the minimal-coupling between mass and gravitational waves in the TT gauge, the interaction Hamiltonian reads
\begin{equation}\label{HintTT}
    \hat{H}_{int}=-\frac{1}{2} \int d\mathbf{x}\ \hat{h}_{ij}^{TT}(\mathbf{x}) \hat{T}^{ij}(\mathbf{x})\, ,
\end{equation}
where $\hat{T}_{ij}(\mathbf{x})$ represents the energy-momentum tensor of matter. In momentum space, the interaction term can be expanded as
\begin{equation}\label{Coupling}
    \hat{H}_{int}=-\frac{\mathcal{A}}{2}\sum_\lambda \int \frac{d\mathbf{k}}{(2\pi)^3} \frac{1}{\sqrt{2\omega_k}} \, {\rm e}^{ij}_\lambda (\mathbf{n}) \hat{g}_{\mathbf{k},\lambda}T_{ij}(\mathbf{k})+ {\rm H.c.} \, .
\end{equation}

Let us consider a massive particle that is prepared as a superposition state of one-dimensional harmonic motion by the Stern-Gerlach apparatus\cite{Folman2013,Amit_2019,Margalit:2021,Wan16_GM,WanPRA16_GM,Pedernales20_GM,Marshman:2021wyk}. The spatial trajectory $\mathbf{x}_c(t)=\{x_c(t),y_c(t),z_c(t)\}$ is modeled by
\begin{equation}\label{x_p}
    x_c^{\pm}(t)=\pm X_0(1-{\rm cos}(\Omega t)) \, , \quad y_c^{\pm}=z_c^{\pm}=0 \, ,
\end{equation}
where $\Omega$ denotes the oscillation frequency of the particle, the constant $X_0$ is the amplitude of the oscillator, and the subscripts "+" and "-" label the left and right trajectory, respectively. The superposition of the mass generates the superposition of the quantum state of gravitons in a displaced vacuum. In the non-relativistic case, the energy-momentum tensor is given by
\begin{equation}\label{Tij}
    T_{ij}(\mathbf{k})= M(\dot{\mathbf{x}}_c)_i (\dot{\mathbf{x}}_c)_j \exp \left( -i\mathbf{k}\cdot{\mathbf{x}}_c - \frac{1}{2}\sigma^2 |\mathbf{k}|^2  \right)\, .
\end{equation}

\begin{figure}[ht]
    \centering
\includegraphics[width=0.8\linewidth]{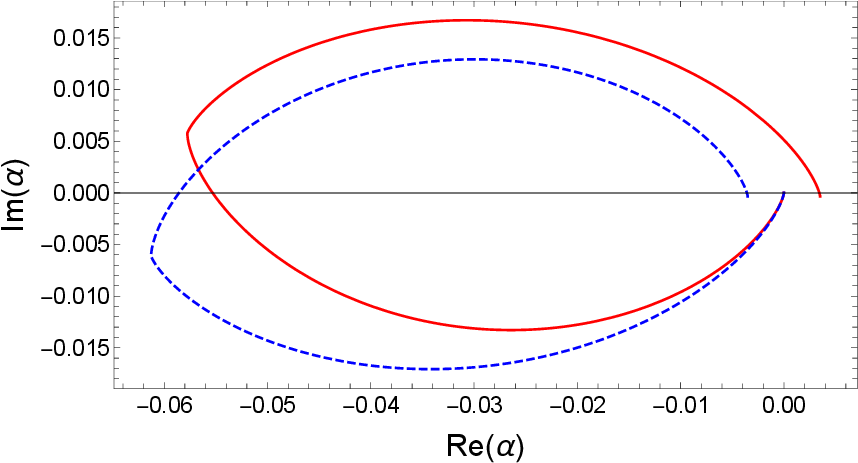}
    \caption{Evolution of the coherent state parameter $\alpha_{\mathbf{k},\lambda}^+$(solid curve) and $\alpha_{\mathbf{k},\lambda}^-$(dashed curve) of a certain graviton mode, where we set $\Omega=1$, $\omega_{\mathbf{k}}=|\mathbf{k}|=1$, $\mathbf{k}=(1,0,0)$, $X_0=0.1$, $\mathcal{C}_{\mathbf{k},\lambda}=1$. The evolution time is from 0 to 1.}
    \label{alpha}
\end{figure}

Here we are working in the interaction picture, where the time-dependent interaction Hamiltonian is
\begin{equation}
\hat{H}^I_{int}= e^{i\hat{H}_0t} \hat{H}_{int} e^{-i\hat{H}_0t} \, .
\end{equation}
We assume that the initial state of the graviton is the vacuum state $\ket{0}$ for all modes with wave vector $\mathbf{k}$ and polarization $\lambda$. 
The evolution operator ${\rm exp}(-i\hat{H}_{int}^I t)$ is equivalent to a displacement operator for each graviton mode, namely
\begin{equation}\label{displacement}
e^{-i\hat{H}_{int}^It} \ket{0}=\prod_{\mathbf{k},\lambda}\, 
e^{i\varphi_{k,\lambda}}  \hat{D}(\alpha_{\mathbf{k,\lambda}}(t))\ket{0}\, ,
\end{equation}
where the displacement is defined by
\begin{equation}
    \hat{D}(\alpha_{\mathbf{k},\lambda}(t))\equiv {\rm exp}(\alpha_{\mathbf{k},\lambda} \hat{g}_{\mathbf{k},\lambda}^\dagger-\alpha_{\mathbf{k},\lambda}^*\hat{g}_{\mathbf{k},\lambda})\, .
\end{equation}
The displacement parameter $\alpha_{\mathbf{k},\lambda}$ and the evolution phase $\varphi_{\mathbf{k},\lambda}$ in Eq.\eqref{displacement} satisfies the differential equations:
\begin{equation}\label{eq:alpha}
\left\{ 
\begin{aligned}
    &i \frac{\partial}{\partial t} \alpha_{\mathbf{k},\lambda}=  \mathcal{C}_{\mathbf{k},\lambda} \sin(\Omega t)^2 e^{i\omega t-i\mathbf{k\cdot \mathbf{x}_p}-\sigma^2\omega^2_{\mathbf{k}}/2} \, , \\
    &-i \frac{\partial}{\partial t} \alpha^*_{\mathbf{k},\lambda}= \mathcal{C}_{\mathbf{k},\lambda} \sin(\Omega t)^2 e^{-i\omega t+i\mathbf{k\cdot \mathbf{x}_p}-\sigma^2\omega^2_{\mathbf{k}}/2} \, ,\\
    & \frac{\partial \varphi_{\mathbf{k},\lambda}}{\partial t}=\frac{i}{2}\left( \frac{\partial\alpha^*}{\partial t}\alpha - \frac{\partial\alpha}{\partial t}\alpha^*\right) \, , \\
    \end{aligned}
\right.
\end{equation}
where we define
\begin{equation}
    \mathcal{C}_{\mathbf{k},\lambda} \equiv \sqrt{\frac{G}{16\pi^5}}\, M\Omega^2X_0^2\, \frac{{\rm e}^{11}_\lambda (\mathbf{n})}{\sqrt{2\omega_{\mathbf{k}}}}\, .
\end{equation}
Therefore, by Eq.\eqref{eq:alpha} and the trajectory Eq.\eqref{x_p}, the parameter $\alpha_{\mathbf{k},\lambda}$ is sovled by
\begin{equation}\label{alpha_klambda}
    \alpha_{\mathbf{k},\lambda}^\pm=-i\mathcal{C}_{\mathbf{k},\lambda}\int_0^t dt'\, \sin(\Omega t')^2 e^{i\omega_k t'\mp ik_x X_0(1-{\rm cos}\Omega t')}\, .
\end{equation}
In Fig.~\ref{alpha}, we show the evolution of the coherent state $\alpha_{{\bf k}, \lambda}$ for specific values set in the caption, for the purpose of illustration.

Thus, the quantum state of the system can be written as 
\begin{equation}\label{gravitonsuperposition}
    \ket{\Psi(t)}=\frac{1}{\sqrt{\mathcal{N}}}(\ket{\{\alpha_+(t)\}}\ket{\psi_L}+\ket{\{\alpha_-(t)\}}\ket{\psi_R})\, ,
\end{equation}
where $\mathcal{N}$ is the normalization factor, $\ket{\psi_{L,R}}$ represents the quantum state of the mass. $\ket{\{\alpha_\pm\}}$ represents the multimode coherent state of gravitons, which is the product over the coherent states of all modes
\begin{equation}
    \ket{\{\alpha_\pm\}}= \bigotimes_{\mathbf{k},\lambda} \ket{\alpha^\pm_{\mathbf{k},\lambda}}\, .
\end{equation}
The overlap between the superposition of the displaced graviton state is evaluated by
\begin{widetext}
\begin{equation}\label{GWoverlap}
    C\equiv\langle \{\alpha_+\}|\{\alpha_-\} \rangle 
    =\exp\left\{-\int d \mathbf{k} \sum_\lambda \frac{1}{2}\left[|\alpha^+_{\mathbf{k},\lambda}|^2 + |\alpha^-_{\mathbf{k},\lambda}|^2 - 2(\alpha^+_{\mathbf{k},\lambda})^*
    \alpha^-_{\mathbf{k},\lambda}\right]\right\} \, .
\end{equation}
\end{widetext}
From Eq.\eqref{alpha_klambda}, we have
\begin{widetext}
\begin{align}
    \int d \mathbf{k}\sum_\lambda  |\alpha^+_{\mathbf{k},\lambda}|^2 =& \mathcal{C}^2 \int d\mathbf{k} \int_0^t dt_1' dt_2' \frac{\sin(\Omega t_1')^2\sin(\Omega t_2')^2}{2\omega_\mathbf{k}} [\sum_\lambda e_{\lambda}^{11}(\mathbf{n})e_{\lambda}^{11}(\mathbf{n}) ]\nonumber\\
    &\times \exp[i\omega_\mathbf{k}(t_1'-t_2')+ikX_0\cos\theta(\cos\Omega t'_1-\cos\Omega t'_2)-\sigma^2\omega_{\mathbf{k}}^2] \nonumber\\
    =&\mathcal{C}^2 \int_0^t dt_1' dt_2' \int_0^\infty dk \int_0^\pi d\theta \int_0^{2\pi} d\phi \frac{k^2(\sin \theta)^3}{2\omega_\mathbf{k}} \sin(\Omega t_1')^2\sin(\Omega t_2')^2  \nonumber\\
    & \times \exp[i\omega_\mathbf{k}(t_1'-t_2')+ikX_0\cos\theta(\cos\Omega t_1'-\cos\Omega t_2')-\sigma^2\omega_{\mathbf{k}}^2] \nonumber\\
    =& 4\pi\mathcal{C}^2 \int_0^t dt_1' dt_2'  \int_0^\infty dk\ \frac{\sin(kX_t)-kX_t\cos(kX_t)}{k^2X_t^3} \sin(\Omega t_1')^2\sin(\Omega t_2')^2 e^{ik(t_1'-t_2')-\sigma^2k^2}\nonumber\\
    =& 8\pi\mathcal{C}^2 \int_0^t dt_1'\int_0^{t'_1} dt_2'  \int_0^\infty dk\ \frac{\sin(kX_t)-kX_t\cos(kX_t)}{k^2X_t^3} \cos[k(t_1'-t_2')] \sin(\Omega t_1')^2\sin(\Omega t_2')^2 e^{-\sigma^2k^2} \, ,
\end{align}
\end{widetext}
where $\theta$ is the angle between $\mathbf{k}$ and $\hat{x}-$axis and we have defined
\begin{equation}
    \mathcal{C}^2\equiv \frac{GM^2}{16\pi^5}\Omega^4X_0^4 \, ,
\end{equation}
\begin{equation}
    X_t\equiv X_0(\cos\Omega t'_1-\cos\Omega t'_2)\, ,
\end{equation}
and we have used the formula\cite{bose2022infrared}
\begin{equation}
    \sum_\lambda e_{\lambda}^{11}(\mathbf{n})e_{\lambda}^{11}(\mathbf{n})= \sin^2\theta \, .
\end{equation}

Similarly, we have
\begin{equation}
    \int d \mathbf{k}\sum_\lambda |\alpha^-_{\mathbf{k},\lambda}|^2 = \int d \mathbf{k} \sum_\lambda|\alpha^+_{\mathbf{k},\lambda}|^2\, .
\end{equation}
and
\begin{align}
    &\int d \mathbf{k}\sum_\lambda (\alpha^+_{\mathbf{k},\lambda})^*
    \alpha^-_{\mathbf{k},\lambda} = \int d \mathbf{k} \sum_\lambda(\alpha^-_{\mathbf{k},\lambda})^*
    \alpha^+_{\mathbf{k},\lambda} \nonumber\\
    &=8\pi\mathcal{C}^2 \int_0^t dt_1'\int_0^{t'_1} dt_2'  \int_0^\infty dk\ \frac{\sin(kX'_t)-kX'_t\cos(kX'_t)}{k^2(X'_t)^3} \nonumber\\
    &\quad\times\cos[k(t_1'-t_2')]\sin(\Omega t_1')^2\sin(\Omega t_2')^2e^{-\sigma^2k^2} \, ,
\end{align}
where we define
\begin{equation}
    X'_t\equiv X_0(2-\cos\Omega t'_1-\cos\Omega t'_2)\,.
\end{equation}

Therefore, the time evolution of the overlap \eqref{GWoverlap} is determined by the integral
\begin{widetext}
\begin{equation}\label{overlap}
    C (t) =\exp\left[-\frac{GM^2}{2\pi^4}\Omega^4X_0^4\int_0^t dt_1' \int_0^{t_1'} dt_2'\int_0^\infty d k\ K(k,t_1',t_2') \cos(k(t_1'-t_2')) \sin(\Omega t_1')^2\sin(\Omega t_2')^2 e^{-\sigma^2k^2} \right]\, .
\end{equation}
\end{widetext}
where the integral kernel $K(k,t_1',t_2')$ is defined by
\begin{align}\label{kernel}
    K(k,t_1',t_2')=&\frac{\sin(kX_t)-kX_t\cos(kX_t)}{k^2X_t^3}\nonumber \\
    &-\frac{\sin(kX_t')-kX_t'\cos(kX_t')}{k^2(X'_t)^3}\, .
\end{align}

Here, we focus on a special case in which the amplitude of the harmonic oscillation is much smaller than the wavelength of the gravitational waves, namely $kX_0\ll1$. In this case, the integral kernel \eqref{kernel} is 
\begin{equation}
    K(k,t_1',t_2')\approx \frac{k^3}{30}[(X_t')^2-X_t^2]\, .
\end{equation}
Therefore, we can compute the overlap \eqref{overlap} between the coherent states of the gravitons when the matter wave trajectories are closed, namely $t=2\pi/\Omega$, 
\begin{equation}
    C(t=2\pi/\Omega) = \exp\left[ - a GM^2 \Omega^6 X_0^6 \right]\, ,
\end{equation}
where the coefficient $a\approx 0.0126$ and we take the limit $\sigma\rightarrow 0$ for the calculation of the overlap. As we can see, for a large superposition size $\sim X_0$ and mass $\sim M$, the overlap between the coherent states of graviton or the spin contrast of the interferometry decreases; hence, the matter-graviton quantum state approaches maximum entanglement.


\bibliography{reference}

\end{document}